\newif\ifnjp
\njpfalse

\newif\ifquantum
\quantumtrue

\ifnjp
\documentclass[12pt]{iopart} 
\expandafter\let\csname equation*\endcsname\relax
\expandafter\let\csname endequation*\endcsname\relax
\usepackage{amsmath}
\usepackage{iopams}
\else
\ifquantum
\documentclass[a4paper,twocolumn,superscriptaddress,10pt,accepted=2018-05-16]{quantumarticle}
\usepackage[numbers,sort&compress]{natbib}
\usepackage{amsmath}
\usepackage{amssymb} 
\newcommand{\tr}{{\rm Tr}} 
\else
\documentclass[10pt,pra,superscriptaddress,amsmath,amssymb,twocolumn]{revtex4-1}
\usepackage{amssymb} 
\newcommand{\tr}{{\rm Tr}} 
\fi
\fi

\pdfoutput=1

\usepackage{url}

\usepackage{amsthm}
\usepackage{graphicx}
\usepackage[all]{xy}
\usepackage[usenames,dvipsnames]{color}

\newcommand{\mc}[1]{\mathcal #1}

\newcommand{\ave}[1]{\langle #1 \rangle}
\newcommand{\bra}[1]{\langle #1 |}
\newcommand{\ket}[1]{| #1 \rangle}
\newcommand{\braket}[2]{\langle #1 | #2 \rangle}

\newcommand{\proj}[1]{\ket{#1}\bra{#1}}
\newcommand{\one}{{\bf 1}}
\newcommand{\re}{{\rm Re}\,}
\newcommand{\im}{{\rm Im}\,}

\newcommand{\dist}[1]{\|{#1}\|_{\mc E,\rho}^2}

\theoremstyle{plain}

\theoremstyle{definition}

\begin{document}

\title{Coarse-grained distinguishability of field interactions}
\author{C\'edric B\'eny}
\address{Institut f\"ur Theoretische Physik, Leibniz Universit\"at Hannover, Appelstra{\ss}e 2, 30167 Hannover, Germany}
\address{Department of Applied Mathematics, Hanyang University (ERICA), 55 Hanyangdaehak-ro, Ansan, Gyeonggi-do, 426-791, Korea.}

\date{\today}
 
\begin{abstract}
Information-theoretical quantities such as statistical distinguishability typically result from optimisations over all conceivable observables. Physical theories, however, are not generally considered valid for all mathematically allowed measurements. For instance, quantum field theories are not meant to be correct or even consistent at arbitrarily small lengthscales. A general way of limiting such an optimisation to certain observables is to first coarse-grain the states by a quantum channel. We show how to calculate contractive quantum information metrics on coarse-grained equilibrium states of free bosonic systems (Gaussian states), in directions generated by arbitrary perturbations of the Hamiltonian. As an example, we study the Klein-Gordon field. If the phase-space resolution is coarse compared to $\hbar$, the various metrics become equal and the calculations simplify. In that context, we compute the scale dependence of the distinguishability of the quartic interaction.
\end{abstract}

\maketitle

The application of tools from quantum information theory (QIT) to QFT is not an entirely straightforward matter. While standard physical applications of a theory only requires that one be able to compute expectation values of certain specific observables on specific states, QIT often requires optimisations over all possible states or observables. Hence, it requires a very detailed understanding of the operational domain of validity of the theory under study. 

The main formalism for QFT comes with several unique features which can potentially complicate such analysis, such as its lack of a Hilbert space or algebraic formulation, or the unavoidable use of divergent asymptotic series. 
Moreover, most interacting QFTs are not meant to be valid in the continuum. Instead, they are defined only relative to an unphysical {\em ultraviolet regulator} which can be thought of as mimicking an underlying discrete space. The process of renormalisation consists in running the parameters of the theory as function of the regulator in such a way that predictions are independent of it. But even for the simplest interacting QFTs such as quantum electrodynamics, this game fails below a certain finite lengthscale. 

Certainly, in an optimisation involving in principle all observables, one should avoid those for which the theory's predictions are not trusted. A simple way to deal with this issue is to add a cutoff on all momentum integrals involved in calculations, which usually amounts to ``tracing-out'' the high momentum modes~\cite{Gu2010,Miyaji2015,Balasubramanian2015}. Here, we want to examine this question more carefully, by explicitly considering a physical (experimental) limit on the resolutions (spatial or otherwise) of the accessible observables.

A very general way of ``coarse-graining'' a quantum system consists in applying a quantum channel on its density matrix. A quantum channel is a linear map on operators which maps density matrices to density matrices. In the absence of a well defined density matrix, the channel can also be defined in the Heisenberg picture, where it can be used, for instance, map ideal observables (such as field operators) to realistic ones characterised by finite resolutions (such as smeared field operators). Below we introduce a specific class of channels which does just that, but also introduces a finite resolution on the measurement of local field values (quadratures). While the finite spatial resolution makes high momenta effectively unobservable, a low field value resolution hides high order polynomials in the field~\cite{Beny2013,Beny2015a}.

The quantity that we want to compute is the statistical distinguishability between two nearby quantum states after being subjected to a given quantum channel. Specifically, we want to compute the extend to which the ground states (or Gibbs states) of two Hamiltonians, $H + \epsilon V_1$ and $H + \epsilon V_2$ are effectively distinguishable in terms of finite-resolutions measurement, where the limited resolution is characterized by a quantum channel. For simplicity the calculation is to be performed to lowest order in $\epsilon$, with $H$ quadratic in canonical variables (although perturbation theory could be used to compute higher orders). 

The concept of statistical distinguishability refers to a class of quantities from which many other information-theoretical quantities can be derived. They have been used for instance for the characterisation of quantum phase transitions~\cite{Zanardi2007,Gu2010}, and in a recently proposed approach to renormalisation~\cite{Beny2013,Beny2015a}. In addition, integrating such quantity along Hamiltonian paths can be used to obtain bounds on ground-state entanglement, as discussed in the outlook (Section~\ref{outlook}). The type of distinguishability metrics considered here have also been connected to the emergent geometry in relation to renormalization and the AdS/CFT correspondance~\cite{Ryu2006}.

The paper is organized as follows.
In Section~\ref{metrics} we introduce distinguishability metrics on density matrices. In Section~\ref{distinguishability}, we derive a general formula for computing how a perturbation of the Hamiltonian affects the effective distinguishability of its ground state, or Gibbs states (Eq.~\eqref{norm}). 

Both sections are presented in terms of density matrices for general quantum systems, with no specific reference to quantum field theory. In order to apply these to QFT, we use a formalism based on characteristic functions typically used in quantum optics. This is introduced in section~\ref{gaussian}.

In Section~\ref{distinggeneral}, the formula for the distinguishability of Hamiltonian perturbations derived in Section~\ref{distinguishability} is specialized to the case of Gaussian states and Gaussian channels introduced in Section~\ref{gaussian}. 

Those results are then applied to the Klein-Gordon field in Section~\ref{kleingordon}, yielding results which are analyzed in Section~\ref{applications}. Specifically, we study the qualitative influence of the resolution parameters on the asymptotic dependence of distinguishability metrics on {\em scale} (which we define as the minimal spatial resolution).

\section{Quantum information metrics}
\label{metrics}

We focus on the calculation of quantum information metrics which directly generalise the classical Fisher information metric. These are Riemannian metrics on the manifold of mixed states. Infinitesimally, these metrics measure {\em statistical distinguishability} between states, such as in the context of hypothesis testing or parameter estimation. 

In a C$^*$-algebraic framework, the main stage is the algebra $\mathcal A$ generalising the set of linear operators on a Hilbert space. A state is defined as a positive function $\rho$ which assigns a probability $\rho(A)$ to any {\em effect} $A$, i.e., any self-adjoint element $A \in \mathcal A$ with spectrum between $0$ and $1$. In finite dimension $\rho$ can be represented by a density matrix $\hat \rho$ such that $\rho(A) = \tr(\hat\rho A)$.

A Riemannian metric is defined by a scalar product defined in the tangent space at every point of a manifold. A tangent vector to a state $\rho$ with density matrix $\hat\rho$ can be defined by a traceless self-adjoint operator $X$ (because, for $\epsilon$ small enough, $\hat\rho + \epsilon X$ must be also a density matrix). Hence we write the tangent space at $\hat\rho$ as
\begin{equation}
T_\rho \equiv \{ X^\dagger = X, \tr(X) = 0 \}.
\end{equation}
The positivity constraint does not appear here because we are interested in the bulk of the manifold. We will deal with boundaries (where the density matrix is not full rank) by approaching them from the bulk.

A metric associates to every $\rho$ a positive linear operator $\Omega_\rho^{-1}$ on $T_\rho$ defining the scalar product
\begin{equation}
X,Y \;\; \longmapsto \;\; \tr(X \Omega_\rho^{-1}(Y)).
\end{equation}
The quantity $\epsilon \, \tr(X \Omega_\rho^{-1}(X))^{\frac 1 2}$ is the distance between $\hat\rho$ and $\hat\rho+\epsilon X$ as $\epsilon \rightarrow 0$. 

Of particular interest are the Riemannian metrics which are contractive under the action of any channel (see below). This contractivity is required for the metric to represent any type of information-theoretic quantity such as distinguishability. In fact, classically, the contractivity condition selects a single metric: the Fisher information metric.

In quantum theory, Petz and Sud\'ar~\cite{Petz1996,Petz1996a} showed that these contractive metrics are one-to-one with {\em operator monotone} function $\theta: \mathbb R^+ \rightarrow \mathbb R^+$ such that $\theta(t) = t \theta(t^{-1})$ for all $t>0$. An operator monotone function has the property that, when applied to operators via functional calculus, $\theta(A) \le \theta (B)$ whenever $A < B$ (i.e., $B - A$ is positive). 

The function $\theta$ defines the kernel $\Omega^{-1}_\rho$ via its inverse $\Omega_\rho$ as follows:
\begin{equation}
\Omega_\rho = 
\theta(L_\rho R_\rho^{-1}) R_\rho,
\end{equation}
where $R_\rho(A) := A \hat \rho$ and $L_\rho(A) := \hat \rho A$ for any matrix $A$.

For instance, in classical probability theory, which is equivalent to considering only density matrices diagonal in the same basis (the diagonal elements then are a probability distribution), since superoperators $R_\rho$ and $L_\rho$ commute, we obtain simply $\Omega_\rho = \theta(1) R_\rho$. The number $\theta(1)$ is an unimportant overall factor which we can pick to be $\theta(1) = 1$, leading to
\begin{equation}
\Omega_\rho(A) = A \hat\rho  \quad \quad \text{(classical)},
\end{equation}
which is independent of the function $\theta$. The resulting metric on probability distributions is simply the Fisher information metric. 
Therefore, the contractive metric parametrised by $\theta$ are all possible non-commutative generalisations of the Fisher information metric. 

We will see that it is generally more convenient to work in the {\em cotangent space} at $\rho$, namely the linear dual $T^*_\rho$ of $T_\rho$. It is the set of linear functionals on the real vector space $T_\rho$. These functionals can be characterised by matrices $A^\dagger = A$ through $X \mapsto \tr(AX)$. Moreover, since $\tr(X) = 0$, we are free to choose an additional constraint which we take to be $\tr(A\hat\rho) = 0$. In this manner, we simply have $\tr((\hat\rho + \epsilon X) A) = \epsilon \tr(XA)$, which directly gives an interpretation of the functional associated with $A$ as an {\em observable}. Moreover, this extra condition makes sense in infinite dimension where the trace may not exists. Hence we write the cotangent space as
\begin{equation}
T_\rho^* \equiv \{ A^\dagger = A, \; \rho(A) = 0\}.
\end{equation}
The metric induces also a scalar product on $T_\rho^*$ which is 
\(
\ave{A,B}_\rho = \tr(A \Omega_\rho(B)).
\)
In infinite dimension, we can remove the trace by defining 
\begin{equation}
\Theta_\rho := \theta(L_\rho R_\rho^{-1}),
\end{equation}
so that $R_\rho \Theta_\rho = \Omega_\rho$, leading to the expression
\begin{equation}
\label{metric}
\ave{A,B}_\rho = \rho( A\, \Theta_\rho(B)),
\end{equation}
where we used the fact that $[R_\rho,L_\rho] = 0$. Observe moreover, that if $\rho$ is a thermal state $e^{-\beta H}$, then the superoperator $L_\rho R_\rho^{-1}$ simply yields the imaginary time evolution:
\begin{equation}
L_\rho R_\rho^{-1}(A) = \rho A \rho^{-1} = e^{-\beta H} A e^{\beta H}.
\end{equation}

One can freely move between the tangent (Schr\"odinger) picture and the cotangent (Heisenberg) picture by using the metric kernel $\Omega_\rho^{-1}$. Indeed, contracting the metric with a tangent vector gives a cotangent vector. Hence, if $\Omega_\rho(A) = X$ and $\Omega_\rho(B) =Y$, then
\begin{equation}
\label{duality}
\tr(X \Omega_\rho^{-1}(Y)) = \ave{A,B}_{\rho}.
\end{equation}
 
Particular contractive metrics which appear in the literature are given by differentiating certain measure of distinguishability which have operational interpretations. That is, the geodesic distance matches the corresponding measure to lowest order (in the distance). 

For instance, differentiating the relative entropy yields the Kubo-Mori metrics defined in Equ.~\eqref{kubomori} below. It is of particular importance to us because of its relation to first-order perturbation theory, as explained in the next section. 

Perhaps the most important metric is the Bures metric, defined by $\theta(t) = (1 + t)/2$ because it has several nice features. Its geodesic distance has a closed analytical form as the Bures distance~\cite{Jencova2004}. It also gives a tight bound on the variance of parameter estimation~\cite{Braunstein1994,Petz2002}, and is as such usually called the quantum Fisher information. Moreover, the Bures metric is the smallest of the contractive metrics (normalised by $\theta(1)=1$)~\cite{Petz1996a}.

We will also refer to the ``square-root'' metric defined by $\theta(t) = t^{\tfrac 1 2}$, because it is especially easy to compute.

\section{Coarse-grained distinguishability}
\label{distinguishability}

In what follows, given a linear ``superoperator'' $\mc E$ defined on the algebra of observables, we write $\mc E_*$ for its ``pre-dual'', namely its adjoint with respect to the Hilbert-Schmidt inner product:
\begin{equation}
\rho(\mc E(A)) = \tr(\mc E(A)\hat\rho) = \tr(A \,\mc E_*(\hat\rho))
\end{equation}
for any observable $A$ and any state $\hat\rho$. We are particularly interested in the case where $\mc E$ is a unital completely positive map, that is, $\mc E(\one) = \one$, and $(\mc E \otimes {\rm id_n})(A) \ge 0$ for all $A \ge 0$ and all finite extra dimension $n$. These conditions guarantee that $\rho \circ (\mc E \otimes {\rm id_n})$ (with density matrix $(\mc E_* \otimes {\rm id})(\hat \rho)$) is a valid state whenever $\rho$ is.

If $\mc E$ is such a unital completely positive map, it represents the action of a {\em quantum channel} in the Heisenberg picture, while $\mc E_*$ represents the same transformation in the Schr\"odinger picture. On infinite-dimensional systems, $\mc E$ is always defined, but $\mc E_*$ may not be. 

Let 
\begin{equation}
\hat\rho := \frac 1 Z e^{-\beta H}
\end{equation}
be the thermal state for the Hamiltonian $H$ at inverse temperature $\beta$.
Geometrically, we want to compute the information metrics pulled back to the manifold of Hamiltonians.

Physically, this has the following interpretation. Given the metric $\Omega_\rho^{-1}$ which gives rise to the geodesic distance $d$, and given the channel $\mc E$, and the Hamiltonian $H$, we want to compute the coarse-grained distance
\begin{equation}
d\bigl({ \mc E_*(\hat\rho_1^\lambda), \mc E_*(\hat\rho_2^\lambda) }\bigr)
\end{equation}
to lowest order in $\lambda$
where
\begin{equation}
\hat\rho_i^\lambda = \frac 1 {Z_{i}(\lambda)} e^{-\beta (H + \lambda V_i)}
\end{equation}
are also normalised thermal states. For notational convenience, we assume that 
\begin{equation}
\tr( \hat\rho (V_2-V_1)) = 0.
\end{equation}  

We have
\begin{equation}
d\bigl({ \mc E_*(\hat\rho_1^\lambda), \mc E_*(\hat\rho_2^\lambda) }\bigr) = \lambda \|V_2-V_1\|_{\mc E,\rho} + \mc O(\lambda^2)
\end{equation}
with
\begin{equation}
\label{norm0}
 \|V\|_{\mc E,\rho}^2 := \beta^2 \tr( \mc E_*\Omega_\rho^S(V)\, \Omega_{\rho \circ \mc E}^{-1} \mc E_*\Omega_\rho^S(V) )
\end{equation}
and $\Omega_\rho^S(V)$ is the first-order term in the imaginary time Dyson series:
\begin{equation}
\label{smetric}
\begin{split}
\Omega_\rho^S(V) &:= \int_0^1 {\hat \rho}^{s} V {\hat\rho}^{1-s} ds\\
&= \frac 1 \beta \Bigl({ \int_0^\beta e^{-s H} V e^{s H} ds}\Bigr)\, \hat\rho\\
&= - \frac 1 {\beta Z} \frac{d}{d\lambda} e^{-\beta (H + \lambda V)}|_{\lambda=0}. \\
\end{split}
\end{equation} 

 because $\Omega_\rho^S(\rho) = \one$)
Incidentally, the inverse of the superoperator $\Omega_\rho^S$ also defines one of the monotone metrics: the Kubo-Mori metric, defined from the function
\begin{equation} 
\label{kubomori}
\theta^S(x) = \int_0^1 x^{s} ds.
\end{equation}

Equ.~\eqref{norm0} needs to be reformulated in a way which is tractable and suitable for the Gaussian and QFT formalism. As we have seen in the previous section, the metrics are easier to formulate in the Heisenberg picture (i.e. on the cotangent space). 

For this purpose it proves useful to define the map 
\begin{equation}
\label{rexplicit}
\mc R_\rho := \Omega_{\rho \circ \mc E}^{-1} \mc E_* \Omega_\rho^S,
\end{equation}
which is the linearisation of the diffeomorphism $H \mapsto H' = \log \mc E_*(e^{-\beta H})$, where $H'$ can be thought of as a coarse-grained effective Hamiltonian since $e^{-\beta H'} = \mc E_*(e^{-\beta H})$. Note that the map $H \mapsto H'$ is the renormalisation group transformation introduced in Ref.~\cite{Beny2015c}.
 
This allows us to rewrite the norm in Equ.~\eqref{norm0} as
\begin{equation}
\begin{split}
\|V\|_{\mc E,\rho}^2 
&= \beta^2 \tr( \mc E_*\Omega_\rho^S(V)\, \mc R_\rho (V) )\\
&= \beta^2 \tr( V \,\Omega_\rho^S \mc E \mc R_\rho(V)  )\\
\end{split}
\end{equation}
where we used the cyclicity of the trace and the fact that the metrics are symmetric.

Writing 
\begin{equation}
\ave{A,B}_\rho = \tr(A \Omega_\rho(B))
\end{equation}
and 
\begin{equation}
\ave{A,B}_\rho^S = \tr(A \Omega_\rho^S(B))
\end{equation}
for the cotangent scalar products associated to the two metrics, one can check that $\mc R_\rho$ introduced above is defined by the fact that it is the adjoint of $\mc E$ with respect to these two scalar products as follows
\begin{equation}
\label{recovery}
\ave{A, \mc E(B)}_\rho^S = \ave{\mc R_\rho(A),B}_{\rho \circ \mc E}
\end{equation}
for all cotangent observables $A$, $B$. Because of this, we call $\ave{\cdot,\cdot}_\rho^S$ the {\em source metric}, and $\ave{\cdot,\cdot}_\rho$, which can still be any of the contractive metrics, the {\em target metric}.
This is the expression that we use to compute the effect of $\mc R_\rho$. 
We can then compute the distinguishability via
\begin{equation}
\label{norm}
\|V\|_{\mc E,\rho}^2 = \beta^2 \ave{V , \mc E \mc R_\rho (V) }_\rho^{S}.
\end{equation}

For context, we note that if the both source and target metrics were the square-root metric, then $\mc R_\rho$ would be a quantum channel, namely the {\rm transpose channel}~\cite{Ohya2004}, which is a useful approximate recovery channel in quantum error correction~\cite{Barnum2002,Ng2010}. Classically, as both metrics reduce to the unique Fisher metric, $\mc R_\rho$ implements Bayesian inference from the conditional probabilities defined by $\mc E$, relative to the prior $\rho$~\cite{Beny2013}.

\section{Gaussian states and channels}
\label{gaussian}

The theory of Gaussian states and Gaussian channels can be formulated directly in an infinite-dimensional setting with uncountable number of degrees of freedom using the formalism of CCR (canonical commutation relations) $C^*$-algebras. For simplicity, and clarity towards the intended audience, however, we do not work in full abstract generality. However, we use a formalism which should be relatively straightforward to generalise if needed. 

The beauty of the Gaussian formalism is that it provides a one-to-one mapping between questions about a quantum system, to questions about a corresponding classical system. It is a mathematically rigorous formalisation of the quantisation of free fields. 

\subsection{CCR algebra}

Accordingly, we start by considering a classical phase-space, defined by a {\em real} vector space $V$ equipped with a symplectic product (an anti-symmetric bilinear form) $\sigma(f,g)$, $f,g \in V$. One may want to think of these phase-space points $f,g$ as classical fields.
The form $\sigma(\cdot,\cdot)$ must be non-degenerate in the sense that if $\sigma(f,g) = 0$ for all $f$ then $g = 0$. For simplicity, it will be convenient to assume that $V$ is equipped with a real scalar product $(f,g)$ which is such that we can write $\sigma(f,g) = (f, \Delta g)$ where $\Delta$ is an anti-symmetric real linear operator on $V$.

Below, we will need to consider transformations of $V$ corresponding to imaginary time evolution, which requires that we work in a complexification of $V$, which we call $V^{\mathbb C}$, where the scalar product is extended to a sesquilinear form via
\begin{equation}
(f+ig,f+ig) := (f,f) + (g,g) - i(g,f) + i(f,g).
\end{equation}
The original $V$ lives on as a subset of $V^{\mathbb C}$, and we call $f \in V$ a {\em real} vector of $V^{\mathbb C}$.

A general classical observable is any real function on $V$, but we will focus only on observables which are {\em linear} functions on $V$, because these are the only ones which can be quantized unambiguously. 
An element $f$ of the phase-space $V$ can be mapped to a linear observable $\Phi_f$ defined by $\Phi_f(g) := (f,\Delta g)$. We extend $\Phi$ by linearity to the whole of $V^{\mathbb C}$, i.e., 
\(
\Phi_{f+ig} := \Phi_{f} + i \Phi_g
\)
if $f,g \in V$.

The form $\sigma(\cdot,\cdot)$, or equivalently $\Delta$, defines the Poisson bracket on those linear classical observables via
\begin{equation}
\{\Phi_f, \Phi_g\} = (f,\Delta g) \one,
\end{equation}
for all $f,g$ real, where $\one(f) = 1$ for all $f$. 

The notation $\Phi_f$ we use is there to evoke the fact that $\Phi_f$ should be thought of as a smeared field observable. To make this clear, let us consider an example, where points of $V$ are given by a pair of a field and its canonical conjugate: $f = (\phi,\pi) \in V$. Here, the canonical field observable usually denoted ``$\phi(x)$'' would actually be the map on $V$ which extract the component $\phi(x) \in \mathbb R$ from $f$, namely, the function $f \mapsto \phi(x)$. If the Dirac delta $(\delta_x,0)$ was an element of $V$, we could write $\phi(x) \equiv \Phi((\delta_x,0))$, although typically it is not, which is why we need this somewhat more general formalism.

Those linear observables $\Phi_f$ for $f \in V$ are the ones that can be unambiguously ``quantized'' to operators $\hat\Phi_f$. However, since these are unbounded operators, it is mathematically more convenient to define instead the objects that would correspond to the Weyl operators $W_f = e^{i \hat \Phi_f}$ (also called displacement operators in optics). Indeed, one can define the CCR algebra $\mathcal A(V)$ associated with $V$ as that generated by the elements $W_f$ defined by the relations
$W_f W_g = e^{- \frac i 2 (f,\Delta g)} W_{f+g}$ and $W_f^\dagger = W_{-f}$ for all $f \in V$. 
One can show that the resulting $C^*$-algebra is essentially unique, and the operators $W_f$ are unitary for all $f \in V$. Once the algebra is represented on a Hilbert space then the objects $\hat \Phi_f$ can be also be defined as operators. 

When extended to the complex $V^{\mathbb C}$, these relations become 
\begin{equation}
\label{weyl}
W_f^\dagger W_g = e^{\frac i 2 (f,\Delta g)} W_{g-\overline f}, \quad \text{and} \quad W_f^\dagger = W_{-\overline f},
\end{equation}
for all $f,g \in V^{\mathbb C}$.

If this algebra can be represented as that of bounded operators on a Hilbert space $\mathcal H$, such that the unitary groups $t \mapsto W(tf)$ have generators, then these are unbounded operators $\hat\Phi_f$, such that 
\begin{equation}
W_f = e^{i \hat\Phi_f}, 
\end{equation}
satisfying the commutation relations
\begin{equation}
[\hat\Phi_f,\hat\Phi_g] = i (\overline f, \Delta g) \one.
\end{equation}
Hence, if these operators $\hat\Phi_f$ exist, we can think of them as the quantizations of the classical observables $\Phi_f$. Moreover, like their classical counterparts, they are linear in their argument:
\begin{equation}
\hat\Phi_{a f + b g} = a \hat\Phi_{f} + b \hat\Phi_{g}
\end{equation}
for all $f,g \in V^{\mathbb C}$ and $a,b \in \mathbb C$.
For most calculations, however, we only need to work with the Weyl operators $W_f$.

In the case of ``second quantization'', such as non-relativistic quantum field theory, the classical phase space $V$ is related to the Hilbert space $\mathcal H$ of ``first quantized'' wavefunctions as follows: the real and imaginary components of the wavefunction $\psi$ play the role of canonical conjugate variables. Hence $V$ is just $\mathcal H$ conceived as a real vector space. If $\braket \psi \phi$ denotes the complex scalar product of $\mc H$, we may use on $V$ the real scalar product $(\psi,\phi) := \re \braket \psi \phi$. The symplectic form is given by the linear operator $\Delta = -2i \one$ as
\begin{equation}
\sigma(\psi,\phi) = (\psi,\Delta \phi) = 2\,\im \braket \psi \phi.
\end{equation}
In this example, the complexification $V^{\mathbb C}$ is not equal to $\mc H$ (it has double the dimension). Consequently, we avoid this formalism.

\subsection{Gaussian states}

It can be deduced from Equ.~\eqref{weyl} that the whole algebra of observables $\mathcal A(V)$ is linearly spanned by the Weyl operators $W_f$. Hence, a state $\rho$ is entirely characterised by its value on those. These values are summarized by the state's {\em characteristic function} $f \mapsto \rho(W_f)$.

A {\em Gaussian state} is one whose characteristic function is Gaussian:
\(
\rho(W_f) = e^{-\frac 1 2 (f,A f) + i (f, f_0)}
\)
for all $f \in V$ and some $f_0 \in V$, where $A$ is a symmetric bilinear operator on $V$. When extended by linearity on $V^{\mathbb C}$, we therefore have 
\begin{equation}
A^\dagger = A^T = A = \overline A.
\end{equation} 
Example of Gaussian states are the thermal states of free bosonic field theories. A similar formalism exists for fermionic fields, but we treat only the bosonic case in this paper.
In what follows, we assume $f_0 = 0$ for simplicity, hence
\begin{equation}
\label{gauss}
\rho(W_f) = e^{-\frac 1 2 (\overline f, A f)}
\end{equation}
for all $f \in V^{\mathbb C}$.
This equation together with Equ.~\eqref{weyl} allows one to compute the expectation value of any operator. For instance, the expectation values of products of smeared field operators can be evaluated by successive differentiation of this expression. We find
\begin{equation}
\begin{split}
\ave{\hat\Phi_f^\dagger \hat\Phi_g}_\rho &\equiv \rho(\hat\Phi_f^\dagger \hat\Phi_g) \\
&= \frac{d^2}{dt\,ds} \rho(W_{tf}^\dagger W_{sg})|_{t=s=0} \\
&= (f,(A + \frac i 2\Delta) g),
\end{split}
\end{equation}
Hence the operator $A$ determines (and is determined by) the real part of the 2-point correlations functions. 

The above expression also implies that the complex operator $A + \frac i 2 \Delta$ must be positive,
\begin{equation}
A + \frac i 2 \Delta \ge 0,
\end{equation}
The operator $A + \frac i 2 \Delta$ may be thought of as the Hamiltonian (rather than Lagrangian) version of the thermal propagator.
 Here, we will call $A$ the {\em covariance operator} as is traditional in the Gaussian formalism.

\subsubsection{Classical Gaussian states}
\label{classicalGaussian}

The above formalism is almost identical for classical statistical theories. The only difference is that $\Delta = 0$, so that the field operators $\hat \Phi_f$ are commuting. Indeed, the Poisson brackets plays no direct role at this level. The Gaussian states can then be interpreted as thermal states of free classical fields living in the phase space $V$, for some Hamiltonian related to $A$. Specifically, given the quadratic classical Hamiltonian
\(
f \mapsto \frac 1 2 (f,H f),
\)
we find that the corresponding thermal state at inverse temperature $\beta$ is the Gaussian defined by the covarience operator
\begin{equation}
A = \frac 1 {\beta}\, H^{-1}.
\end{equation}

\subsection{Gaussian channels}

A {\em channel} in quantum theory refers to the most general map from states to states which is consistent with the probabilistic interpretation of the convex combination of states, as well as with the system being part of a larger one. In the Heisenberg pictures, it must be a linear map between algebras of observables that is {\em completely positive}, and which preserves the identity. Classically, these map correspond to all stochastic maps. In quantum theory, channels include unitary transformation, but also maps which add noise to the system, and correspond to the evolution of open quantum systems. 
We will not go here into details of this definition, because we will simply focus on a special class of channels which map Gaussian states to Gaussian states, the so-called {\em Gaussian channels}. 

As with states, channels are characterised entirely by their action on the Weyl operators. We consider channels $\mc E$ with the following action
\begin{equation}
\label{channel}
\mc E(W_f) = W_{Xf} \,e^{-\frac 1 2 (\overline f, Y f)}
\end{equation}
where $X$ and $Y$ are linear on $V$, i.e., real on $V^{\mathbb C}$: $X^T = X^\dagger$ and $Y^T = Y^\dagger$.
We can easily check that this maps Gaussian states to Gaussian states, provided certain conditions on the linear operators $X$ and $Y$. If $\rho_A$ is the Gaussian state defined by the covariance matrix $A$, we obtain using Equ.~\eqref{gauss} that
\begin{equation}
\label{channelcov}
\rho_A(\mc E(W_f)) =\rho_{X^\dagger A X + Y} (W_f).
\end{equation}
From the requirement that $X^\dagger A X + Y + \frac i 2 \Delta \ge 0$ for all $A + \frac i 2 \Delta \ge 0$, we obtain the condition 
\begin{equation}
\label{chancond}
Y - \frac i 2 X^\dagger \Delta X + \frac i 2 \Delta \ge 0.
\end{equation}

We note that classically, with $\Delta = 0$, this simply reduces to the condition $Y \ge 0$.

\section{Distinguishability near Gaussian states}
\label{distinggeneral}

We now want to obtain the adjoint map $\mc R_\rho$ using Equ.~\eqref{recovery} when $\rho$ is a Gaussian. The strategy is to evaluate all components in terms of the Weyl operators $W_f$, or more conveniently, in terms of the functional
\begin{equation}
\label{gen}
G^A_f := 
W_f\, e^{\frac 1 2 (\overline f,A f)},
\end{equation}
Since those operators are not self-adjoint, we need to extend the metric to all operators, making it sesquilinear. Using Equ.~\eqref{metric}, this is
\begin{equation}
\ave{A,B}_\rho^\theta := \rho(A^\dagger \Theta_\rho (B)).
\end{equation}
We will make use of the fact that, assuming $A$ is the covariance matrix of $\rho$,
\begin{equation}
\label{genexp}
\rho((G^A_f)^\dagger G^A_g) = e^{(f, (A + \frac i 2 \Delta) g)},
\end{equation}
for all $f,g \in V^{\mathbb C}$.
This can be computed directly using Equ.~\eqref{weyl} and Equ.~\eqref{gauss}.

Abbreviating 
\begin{equation}
B = X^\dagger A X+Y
\end{equation}
for the covariance matrix of the coarse-grained state $\rho \circ \mc E$ (See Equ.~\eqref{channelcov}), we also have
\begin{equation}
\rho(\mc E( (G^B_f)^\dagger G^B_g )) = e^{(f, (B + \frac i 2 \Delta) g)}.
\end{equation}
In terms of those functionals, Equ.~\eqref{channelcov} becomes simply
\begin{equation}
\label{channeleffect}
\mc E(G^{B}_f) = G^A_{Xf}.
\end{equation}

The particular metric $\Omega_\rho^S$ defined in Equ.~\eqref{smetric}, corresponds to the operator monotone function 
\begin{equation}
\theta^S(x) = \frac{x-1}{\log x} = \int_0^1 x^{s} ds.
\end{equation}
Recall that $\Theta_\rho$ is given by applying $\theta$ (through functional calculus) to the superoperators of imaginary time evolution, namely the transformations $X \mapsto \rho X \rho^{-1} = e^{-\beta H} X e^{\beta H}$. 
For a Gaussian state this generates a group of (complex) canonical transformations which can be represented by linear operators $R_s^A$ (where $A$ stands for the covariance matrix defining $\rho$) on the complexified phase space:
\begin{equation}
e^{-s \beta H} W_f e^{s \beta H} = W_{R_s^A f}.
\end{equation}
Since $\rho$ is invariant under the imaginary time evolution it defines, we have $R_s^T A R_s = A$, hence
\begin{equation}
e^{-s \beta H} G^A_f e^{s \beta H} = G^A_{R_s^A f}
\end{equation}
Moreover, the fact that it is canonical means that it preserves the symplectic form: $(R_s^A)^T \Delta R_s^A = \Delta$. Another important property of these operators is $\overline {R_s^A} = R_{-s}^A$, from which we obtain that
\begin{equation}
\label{invarience}
(A + \frac i 2 \Delta) R_s^A = (R_s^A)^\dagger (A + \frac i 2 \Delta).
\end{equation}
Also, by computing the components of the map $G_f \mapsto  \rho G_f = G_{R^A_\beta f} \rho$ from both expression, we obtain
\begin{equation}
(A + \frac i 2 \Delta) R^A_1 = A - \frac i 2 \Delta. 
\end{equation}

Explicitly, the Kubo-Mori metric is
\begin{equation}
\label{genprod}
\ave{G^A_f,G^A_g}_\rho^S = \int_0^1 \, e^{(f,(A+\frac i 2 \Delta) R^A_{s} g)} ds.
\end{equation}
Then the definition of $\mc R_\rho$ in Equ.~\eqref{recovery} becomes
\begin{equation}
\label{recoverycomps}
\begin{split}
\ave{\mc R_\rho(G^A_f),G^B_g}_{\rho\circ\mc E} &= \ave{G^A_f,\mc E(G^B_g)}_\rho^S
= \ave{G^A_f,G^A_{Xg}}_\rho^S\\
&= \int_0^1 \, e^{(f,(A+\frac i 2 \Delta) R_s^A X g)} ds.\\
&= \int_0^1 \, e^{(R_s^A f,(A+\frac i 2 \Delta) X g)} ds.\\
\end{split} 
\end{equation}

Equ.~\eqref{recoverycomps} can be used to prove the following. Let $\mc P_A^n$ be the linear space of polynomials obtained by differentiating $G^A(f)$ with respect to its argument $n$ times, i.e.,
\begin{equation}
\mc P_A^n := \Bigl\{{ \frac{\partial^n G^A_{t_1f_1 + \dots + t_nf_n}}{\partial t_1 \dots \partial t_n}\Bigl|_{0} : f_1,\dots,f_n \in V  }\Bigr\}
\end{equation}
where the subscript $0$ indicates the the derivative is evaluated at $t_1=\dots=t_n=0$.
These are polynomials of order $n$ in the field operators. Moreover, the span of $\mc P_A^1$ to $\mc P_A^n$ is the space of all polynomials in the fields of order $n$ (or lower).  

Because the exponent in the last term in Equ.~\eqref{recoverycomps} is linear in both $f$ and $g$, the whole expression is equal to zero whenever one differentiates it with respect to $f$ and $g$ a different number of times (at $f=g=0$). Moreover, Equ.~\eqref{genprod} shows with the same argument that $\mc P_A^n$ is orthogonal to $\mc P_A^m$ in terms of the Kubo-Mori metric at $\rho$ whenever $n\neq m$. 

Similarly, if we assume that the target metric is defined by an operator monotone function of the form $\theta(x) = \int x^s d\mu(s)$, for some measure $\mu$, which includes the Kubo-Mori metric, Bures metric and square-root metric,
then we have explicitly,
\begin{equation}
\label{genprodtarget}
\ave{G^B_f,G^B_g}_{\rho \circ \mc E} = \int \, e^{(R^B_{s} f,(B+\frac i 2 \Delta)  g)} d\mu(s),
\end{equation}
which implies that $\mc P_B^n$ and $\mc P_B^m$ are orthogonal also in terms of that target metric. 
Taken together with the completeness of all these polynomials, this proves that 
\begin{equation}
\mc R_\rho(\mc P_A^n) \subseteq \mc P_B^n.
\end{equation}
Since also, for Equ.~\eqref{channeleffect}, $\mc E(\mc P_B^n)\subseteq \mc P_A^n$, we obtain that
for any Gaussian channel $\mc E$, and any Gaussian state $\rho$ with covariance operator $A$, 
\begin{equation}
\mc E \mc R_\rho(\mc P_A^n) \subseteq \mc P_A^n.
\end{equation}
Moreover, since $\mc P_A^n$ is orthogonal to $\mc P_A^m$ whenever $n \neq m$ in terms of the Kubo-Mori metric $\ave{\cdot,\cdot}_\rho^S$, then
\begin{equation}
\ave{V_n,\mc E \mc R_\rho(V_m)}_\rho^S = 0
\end{equation}
for any $V_n \in \mc P_A^n$ with $n \neq m$. This implies that we can restrict the problem of computing the components of the linear map $\mc R_\rho$ to each subspace $\mc P_A^n$ independently.

With some extra assumption on $\rho$ and the channel, the same argument can yield a more detailed result which will be useful below.
Suppose that $X$, $Y$, $A$ and $\Delta$ are all jointly block diagonal for some decomposition $V = \bigoplus_k V_k$ of the classical phase space (assuming that $k$ is an integer for convenience).
Let us define the operator spaces
\begin{equation}
\mc P_{k_1 < \dots < k_n}^n := \Bigl\{{ \frac{\partial^n G^A_{t_1f_1 + \dots + t_nf_n}}{\partial t_1 \dots \partial t_n}\Bigl|_{0} : f_i \in V_{k_i} \; i=1,\dots,n  }\Bigr\}.
\end{equation}
Then, following the same argument as above, we see that these spaces are all orthogonal to each other in the Kubo-Mori metric at $\rho$, and also that
\begin{equation}
\label{modeindep}
\mc E \mc R_\rho(\mc P_{k_1 < \dots < k_n}^n) \subseteq \mc P_{k_1 < \dots < k_n}^n.
\end{equation}
An example is the Klein-Gordon example studied below, where $k$ index momentum modes, and the spaces $\mc P_{k_1 < \dots < k_n}^n$ are finite-dimensional, allowing for an exact solution for each family of modes. However, for some specific metrics, or with a rather innocuous simplification, the solution can be made much more explicit.

\subsection{Exact solution for square-root metric}
\label{chi2exact}

The quantity $\|V\|_{\mc E,\rho}$, or more generally all components of the coarse-grained metric, can be computed exactly in full generality for the square-root metric, defined by the function
\begin{equation}
\theta(x) = x^\frac 1 2,
\end{equation}
which yields
\begin{equation}
\ave{G^B_{h},G^B_g}_{\rho\circ\mc E} = e^{(R_{1 / 2}^B h,(B+\frac i 2 \Delta) g)}.
\end{equation}
Indeed, we see from comparing this to Equ.~\eqref{recoverycomps} that $\mc R_\rho(G^A_f) = \int_{0}^{1} G^B_{h(s)} ds$ provided that $h(s)$ is such that for all $0 \le s \le 1$ and for all $f, g \in V$,
\begin{equation}
(R_{1/2}^B h(s),(B+\frac i 2 \Delta) g) = (R_{s}^A f,(A+\frac i 2 \Delta) X g),
\end{equation}
namely
\(
h(s) = J_s f
\)
where
\begin{equation}
J_s := R_{-1/2}^B (B+\frac i 2 \Delta)^{-1} X^\dagger (A+\frac i 2 \Delta) R_{s}^A.
\end{equation}
It follows that
\begin{equation}
\mc E \mc R_\rho(G^A_f) = \int_0^1 \mc E(G^B_{J_s f}) ds = \int_0^1 G^A_{X J_s f} ds,
\end{equation}
and
\begin{equation}
\label{chi2gen}
\begin{split}
\ave{G^A_f,\mc E \mc R_\rho(G^A_g)}_\rho^S &= \int_0^1 ds \ave{G^A_f,G^A_{XJ_s g}}_\rho^S\\
&= \int_0^1 ds\,dt\, e^{R_t^A  f,(A+\frac i 2\Delta) X J_s g)} \\
&= \int_0^1 ds\,dt \,e^{(f, (R_t^A)^\dagger  P R_s^A g)} \\
\end{split}
\end{equation}
where
\begin{equation}
\label{chi2P}
P = (A+\frac i 2\Delta) X R_{-1/2}^B (B+\frac i 2 \Delta)^{-1} X^\dagger (A+\frac i 2 \Delta).
\end{equation}
Note that $P^\dagger = P$ because 
\begin{equation}
R_{-1/2}^B (B+\frac i 2 \Delta)^{-1} = R_{-1/4}^B (B+\frac i 2 \Delta)^{-1} (R_{-1/4}^B)^\dagger.
\end{equation}

One can then compute $\|V\|^2_{\mc E,\rho}$ for any operator $V$ by differentiation of the last expression in Equ.~\eqref{chi2gen} with respect to $f$ and $g$. 
In general, one can write the operator $V$ as
\begin{equation}
V = D^V_f G^A_f
\end{equation}
where $D^V_f$ is a differential operator with respect to the observable $f$ (we give some explicit examples below). 
Using Equ.~\eqref{norm0}, this yields
\begin{equation}
\label{genexample}
\|V\|^2_{\mc E,\rho} = \beta^2 \int_0^1  D^V_f D^V_g \,e^{(f, P(s,t) g)} \, ds \,dt,
\end{equation}
where
\begin{equation}
P(s,t) = (R_t^A)^\dagger P R_s^A.
\end{equation}
The result from the action of these differential operators on the exponential generator can be expressed in terms of familiar Feynman diagrams, with the propagator given by relevant components of $P(s,t)$. 

For instance, a differentiation of order $n$ yields a Feynman diagram with $n$ edges in total, which represents a contraction of the operator
\begin{equation}
\label{whattocompute}
\int_0^1 P(s,t) \otimes \dots \otimes P(s,t) \, ds\,dt.
\end{equation}

\subsection{Exact solution for the classical Fisher metric}
\label{classicalsol}

All quantum contractive metrics reduce to the Fisher information metric when all operators commute. For our purpose, this implies $\Delta = 0$, as well as $R_s = \one$ since the imaginary time evolution is trivial: $\rho^{-s} X \rho^s = \rho^{-s} \rho^s X = X$ for all $X$. Hence we can use the above result to directly get the classical solution simply by setting $\Delta = 0$ and $R_{\frac 1 2} = \one$ in Equ.~\eqref{chi2P}, yielding
\begin{equation}
\label{fisherP}
\begin{split}
P &= A X B^{-1} X^\dagger A =  A X (X^\dagger A X + Y)^{-1} X^\dagger A \\
&= A (A + (X^\dagger)^{-1} Y X^{-1} )^{-1} A,
\end{split}
\end{equation} 
and
\begin{equation}
\label{fishergen}
\ave{G^A_f,\mc E \mc R_\rho(G^A_g)}_\rho  = e^{(f, P g)},
\end{equation}
where we removed the $S$ label since all metrics used are the same Fisher metric in this case. This result is equivalent to that obtained in Ref.~\cite{Beny2015a}.

\subsection{Approximation for large noise}
\label{noisyapprox}

There are contexts where the channel, defined by $X$ and $Y$, is noisy enough that the coarse-grained propagator is approximately independent of the imaginary time $s$:
\begin{equation}
(B + \frac i 2 \Delta) R_s^B \simeq K_{\rm app}
\end{equation}
for any $s \in [0,1]$. We will see below an example where this is the case. 
In this case, no matter what target metric we use,
\begin{equation}
\ave{G^B_{h},G^B_g}_{\rho\circ\mc E} \simeq e^{(h, K_{\rm app} g)},
\end{equation}
so that we can use the same method as in Section~\ref{chi2exact} to obtain
\begin{equation}
\label{largeYgen} 
\ave{G^A_f,\mc E \mc R_\rho(G^A_g)}_\rho^S \simeq \int_0^1 ds\, dt \,e^{(f, (R^A_t)^\dagger P R^A_s g)},
\end{equation}
with
\begin{equation}
\label{largeYP} 
P = (A+\frac i 2 \Delta) X K_{\rm app}^{-1} X^\dagger (A + \frac i 2 \Delta).
\end{equation}

\section{Example: Klein-Gordon field}
\label{kleingordon}

The classical Klein-Gordon Hamiltonian can be written as
\begin{equation}  
\tilde{\mc H} = \frac 1 2 \int dx \, \bigl( \tilde \pi(x)^2 + \phi(x)(m^2 - \partial_x^2) \tilde \phi(x)  \bigr),
\end{equation}
where $\tilde \pi(x)$ and $\tilde \phi(x)$ are canonical conjugates. 
The canonical change of variable
\begin{align}
\tilde \Phi_k &= \int dx \bigl[{ \cos(kx) \tilde \phi(x) - \omega_k^{-1} \sin(kx) \tilde \pi(x)}\bigr]\\
\tilde \Pi_k &= \int dx \bigl[{ \omega_k \sin(kx) \tilde \phi(x) + \cos(kx) \tilde \pi(x)}\bigr].\\
\end{align}
yields the decoupled form
\begin{equation}
\tilde{\mc H} = \frac 1 2 \int dk ( \tilde \Pi_k^2 + \omega_k^2 \tilde \Phi_k^2  ).
\end{equation}
where $\omega_k = \sqrt{k^2 + m^2}$ and the Poisson bracket is
\begin{equation}
\{\tilde \Phi_k, \tilde \Pi_{k'}\} = \delta(k-k') \one.
\end{equation}

Instead of attempting to quantize this directly, we consider the discretisation 
\begin{equation}
\mc H = \frac 1 2  \sum_k \frac{1}{L^d} ( \tilde \Pi_k^2 + \omega_k^2 \tilde \Phi_k^2  ) = \frac 1 2  \sum_k (\Pi_k^2 + \omega_k^2 \Phi_k^2 ),
\end{equation}
where
\begin{equation}
\Phi_k = L^{-\frac d 2} \tilde \Phi_k \quad\text{and} \quad \Pi_k = L^{-\frac d 2}  \tilde \Pi_k
\end{equation}
satisfy 
\begin{equation}
\{\Phi_k, \Pi_{k}\} = \one.
\end{equation}
The ``infrared regulator'' $L^d$, where $d$ is the dimension of space, can be thought of as a volume. The original theory with continuous momenta $k$ is recovered for $L \rightarrow \infty$.

The discrete Hamiltonian simply represents a discrete set of decoupled harmonic oscillator. Therefore, the corresponding quantum Hamiltonian is
\begin{equation}
H = \sum_k \omega_k a_k^\dagger a_k 
\end{equation}
where the sum is over some discrete set of modes $k$, and $[a_k,a_{k}^\dagger] = \one$ and $[a_k,a_{k'}] = 0$ for $k \neq k'$.
The quantum versions of the observables $\phi_k$ and $\pi_k$, which we denote by the same symbols, are 
\begin{align*}
\Phi_{k} &= \sqrt{1/2\omega}(a_k^\dagger + a_k),\\
\Pi_{k} &= i \sqrt{\omega/2} (a_k^\dagger - a_k).
\end{align*}

Since there is no interaction between modes $k$, the Gibbs state is of the form $\rho = \bigotimes_k \rho_k$. 
In the basis composed of the classical observables $(\Phi_k,\Pi_k)$, the symplectic form is
\begin{equation}
\Delta_k = \begin{pmatrix} 0 & 1 \\ -1 & 0 \end{pmatrix}.
\end{equation}
Using the creation and annihilation operators, one can easily find the expectation values of products of two field operators, and hence the component of the covariance matrix $A_k$ defining the state $\rho_k$:
\begin{equation}
A_k = \tfrac 1 2 \coth \bigl({ \tfrac{\beta \omega}{2} }\bigr) \begin{pmatrix} \tfrac 1 \omega_k & 0  \\  0 & \omega_k \end{pmatrix}.
\end{equation}
One can obtain the components of $R_s^{A_k}$ by solving the imaginary time equations for the harmonic oscillator, which yields
\begin{equation}
\label{kgra}
R_s^{A_k} = \begin{pmatrix} \cosh(\beta \omega_k s) & - i \omega_k \sinh(\beta \omega_k s) \\ i \sinh(\beta \omega_k s)/\omega_k & \cosh(\beta \omega_k s) \end{pmatrix}.
\end{equation}

We could proceed using the real phase space coordinates $\Phi_k$ and $\Pi_k$. However, the coarse-graining channel that we will use takes a simpler form in terms of the complex variables
\begin{align}
\phi_k = \frac 1 2 ( \Phi_k + \Phi_{-k} ) + \frac{i}{2 \omega_k}(\Pi_k - \Pi_{-k}), \\
\pi_k = \frac 1 2 ( \Pi_k + \Pi_{-k} ) - \frac{i \omega_k }{2 }(\Phi_k - \Phi_{-k}). \\
\end{align}
These observables are the standard Fourier modes used in scalar field theory. They are related to the original fields $\tilde \phi(x)$ and $\tilde \pi(x)$ simply through
\begin{align}
\tilde \phi(x) &= \lim_{L \rightarrow \infty} L^{\frac d 2}\sum_k e^{ikx} \phi_k,\\
\tilde \pi(x) &= \lim_{L \rightarrow \infty} L^{\frac d 2}\sum_k e^{ikx} \pi_k.\\
\end{align}

Together with $(\phi_{-k},\pi_{-k}) = (\overline \phi_k,\overline \pi_k)$, this is just a complex change of coordinate on the four-dimensional subspace of $V^{\mathbb C}$ corresponding to modes $k$ and $-k$. Recall that $\Delta$ and $A$ both define sesquilinear forms on $V^{\mathbb C}$. Therefore, if we denote the components of this coordinate change by the four-by-four matrix $P$, $A_k \oplus A_{-k}$ transforms to $P^\dagger(A_k \oplus A_{-k})P$ in the new coordinate system. One can check by direct calculation that they have the exact same form as before: $P^\dagger(A_k \oplus A_{-k})P = A_k \oplus A_{-k}$ and $P^\dagger(\Delta_k \oplus \Delta_{-k})P = \Delta_k \oplus \Delta_{-k}$. 

From now on, by ``mode $k$'' we mean either the subspace of phase space spanned by $(\phi_k,\pi_k)$, or the corresponding subsystem in the quantum theory. 

In order to proceed further, we need to fix a channel $\mc E$ via the operators $X$ and $Y$. We use the linear operator $X$ defined by
\begin{align}
X \phi_k &= e^{-\frac 1 2 k^2 \sigma^2} \phi_k,\\
X \pi_k &= e^{-\frac 1 2 k^2 \sigma^2} \pi_k,\\
\end{align}
The operator $Y$ defines a sesquilinear form, which we take to be block-diagonal $Y = \bigotimes_{k \in M} Y_k$ where $Y_k$ can be represented by the matrix
\begin{equation}
Y_k 
= \begin{pmatrix} y_\phi^2 & 0 \\ 0 & y_\pi^2 \end{pmatrix}.
\end{equation}
This is a variation of the channel used in Ref.~\cite{Beny2015a}, but this operator $Y$ is different as it couples the real modes $k$ and $-k$ when expressed in terms of the coordinates $(\phi_k,\pi_k)$. The parameter $\sigma$ characterises the maximal precision at which space is resolved by the observer, and the values $L^{\tfrac d 2} y_\phi$ and $L^{\tfrac d 2} y_\pi$ characterise the precision at which the field and canonical field values $\tilde \phi(x)$ and $\tilde \pi(x)$ respectively are resolved at each point in space. 

But recall that $X$ and $Y$ must satisfy Equ.~\eqref{chancond} in order for $\rho \circ \mc E$ to be a valid state, or, said differently, for $\mc E(A)$ to be a positive effect whenever $A$ is. On mode $k$, the equation reduces to
\begin{equation}
Y_k + (1 - e^{-k^2 \sigma^2}) \frac i 2 \Delta \ge 0.
\end{equation}
For $k \ll 1/\sigma$, this just implies that $Y_k \ge 0$. But for $k \gg 1/\sigma$, this yields the non-trivial relation 
\begin{equation}
\label{uncertainty}
y_\phi^2 y_\pi^2 \ge 1.
\end{equation}
This can be understood intuitively by looking at the covariance matrix 
\begin{equation}
B_k = X^\dagger_k A_k X_k + Y_k
\end{equation}
which defines the state $\mc E_k(\rho_k)$. When $X_k \simeq 0$, the covariance matrix is just $Y_k$, hence it must satisfy the Heisenberg uncertainty relation given by Equ.~\eqref{uncertainty}.

It is possible to write $B_k$ in the same form as $A_k$, but in terms of redefined frequencies and temperatures:
\begin{equation}
B_k = \tfrac 1 2 \coth \bigl({ \tfrac{\beta_k' \omega_k'}{2} }\bigr) \begin{pmatrix} \tfrac 1 {\omega_k'} & 0  \\  0 & \omega_k' \end{pmatrix}.
\end{equation}
Let us define
\begin{align}
\label{uk}
u_k^2 &:= y_\phi^2 + \omega_k^{-1} \tfrac 1 2 \coth \bigl( \tfrac{\beta_k \omega_k}2 \bigr) e^{-k^2 \sigma^2},\\
\label{vk}
v_k^2 &:= y_\pi^2 + \omega_k \tfrac 1 2 \coth \bigl( \tfrac{\beta_k \omega_k}2 \bigr) e^{-k^2 \sigma^2}.
\end{align}
Then we have
\begin{equation}  
\label{cgomega}
\omega_k'= \frac{v_k}{u_k}
\end{equation}
and
\begin{equation}  
\label{cgbetaexact}
\beta_k'= \frac{u_k}{v_k} 2 \coth^{-1}(2 u_k v_k).
\end{equation}
Moreover, assuming $y_\phi y_\pi \gg 1$, which implies $u_k v_k \gg 1$ for all $k$, we obtain
\begin{equation}
\label{cgbeta}
\beta_k' \simeq \frac{1}{v_k^2}.
\end{equation}
With these definitions, $R_s^{B_k}$ is given simply by substituting $\omega_k \rightarrow \omega_k'$ and $\beta \rightarrow \beta_k'$ in Equ.~\eqref{kgra}.

But since then $\beta_k' \omega_k' \simeq \frac 1 {\sqrt{ u_k v_k }} \ll 1$, each mode $k$ is in the high temperature limit where all quantum metrics reduce to the classical one. Indeed, we obtain simply
\begin{equation}
R_s^{B} \simeq \one + \mc O( y_\phi y_\pi  )^{-1}
\end{equation}
for all $s$. This puts us in the situation described in Section~\ref{noisyapprox} with
\begin{equation}
\label{thekapp}
K_{\rm app} = \bigoplus_k \begin{pmatrix} u_k^2   & 0 \\ 0 & v_k^2  \end{pmatrix}.
\end{equation}

As an example let's compute the distinguishability of $\phi_k$ and $\pi_k$, which are given by the first order derivatives of $G_f^A$. Differentiating Equ.~\eqref{largeYgen} once with respect to $f$ and $g$, we obtain that 
\begin{equation}
\begin{split} 
M_k &:= \begin{pmatrix}
\ave{\phi_k,\mc E \mc R_\rho (\phi_k)}_\rho^S & \ave{\phi_k,\mc E \mc R_\rho (\pi_k)}_\rho^S\\
\ave{\pi_k,\mc E \mc R_\rho (\phi_k)}_\rho^S & \ave{\pi_k,\mc E \mc R_\rho (\pi_k)}_\rho^S\\
\end{pmatrix}\\
&=
K_S^\dagger \begin{pmatrix} u_k^{-2}  & 0 \\ 0 & v_k^{-2}  \end{pmatrix} K_S e^{- k^2 \sigma^2}
\end{split}
\end{equation}
where
\begin{equation}
K_S = \int_0^1 (A_{k} + \tfrac i 2 \Delta) R_s^{A_{k}}  ds
= \frac 1 \beta
\begin{pmatrix}
\frac 1 {\omega_k^2} & 0\\ 0 & 1
\end{pmatrix}.
\end{equation}
Hence we obtain from the diagonal components of $M_k$ that, with $y_\phi y_\pi \gg 1$,
\begin{align}
\dist{\phi_k} \simeq \frac {e^{-k^2\sigma^2}} { \omega_k^4 u_k^2} \quad \text{and}\quad
\dist{\pi_k} \simeq \frac {e^{-k^2\sigma^2}} { v_k^2}.
\end{align}

For the calculations involved higher order polynomials, we avoid doing the integrals over $t$ and $s$ by considering only the zero temperature limit $\beta \rightarrow \infty$.
Recall that, at finite temperature, 
\begin{equation}
\Omega^S_\rho(V) = - \beta^{-1} \frac{d}{dt} \frac{e^{-\beta (H + t V)}}Z |_{t=0},
\end{equation}
for $V$ self-adjoint and of zero expectation value. 
At zero temperature, the state $e^{-\beta (H + t V)}/Z |_{t=0}$ should become the projector on the ground state $\ket{\Omega_t}$ of the quadratic Hamiltonian $H$ perturbed by $t V$. Hence, if we write $\ket{\Omega_t} = \ket {\Omega_0} + t \ket {\Omega_{1}} + \mc O(t^2)$, we have
\begin{equation}
\begin{split}
\lim_{\beta \rightarrow \infty} \beta \ave{W,V}^S_\rho &= - \frac{d}{dt} \tr(W \proj{\Omega_t})|_{t=0} \\
&= - \bra{\Omega_0} W \ket{\Omega_1} - \bra{\Omega_1} W \ket{\Omega_0}.
\end{split}
\end{equation}
Moreover, from perturbation theory, 
\begin{equation}
 \ket{\Omega_1} = -H^{-1} V \ket{\Omega_0},
\end{equation}
where $H$ is the Hamiltonian, shifted so that the ground state energy is zero, and $H^{-1}$ is defined to be zero on the ground state.
Hence, we can rewrite the Kubo-Mori metric at zero temperature, now for the complexified version, as 
\begin{equation}
\label{metricatzero}
\begin{split}
\lim_{\beta \rightarrow \infty} \beta &\ave{G_f^A,G_g^A}^S_\rho \\
&= \rho( (G_f^A)^\dagger H^{-1} G_g^A )+ \rho( G_g^A  H^{-1} (G_f^A)^\dagger).
\end{split}
\end{equation}

For our example, using the Hamiltonian,
\begin{equation}
H =  \sum_k \,\omega_k a_k^\dagger a_k,
\end{equation}
we can obtain the components of the metric in the sector of distinct
modes $k_1, \dots, k_n$ at $\beta = \infty$. From differentiating Equ.~\eqref{metricatzero}, using Equ.~\eqref{genexp} and Equ.~\eqref{genprod}, we obtain that 
\begin{equation}
\label{zerotempapprox}
\begin{split}
&\lim_{\beta \rightarrow \infty} \beta \int_0^1 \bigotimes_{j=1}^n (A_{k_j} + \frac i 2 \Delta) R^{A_{k_j}}_s \\
& \quad \quad =  \frac{1}{\sum_j \omega_{k_j}} \left[{ \bigotimes_{j=1}^n (A_{k_j}^\infty + \frac i 2 \Delta) +  \bigotimes_{j=1}^n (A_{k_j}^\infty - \frac i 2 \Delta)}\right],
\end{split}
\end{equation}
where $A_k^\infty = \lim_{\beta \rightarrow \infty} A_k = \frac 1 2 \begin{pmatrix} \frac 1 {\omega_{k_j}} & 0  \\  0 & \omega_{k_j}\end{pmatrix}$.

We used the fact that
\begin{equation}
\label{derivative}
\begin{split}
&\partial_{t_1}\cdots \partial_{t_n} \partial_{s_1}\cdots \partial_{s_n} e^{\sum_{i,j=1}^n t_i s_j(f_{k_i},P g_{k_j})}|_{t_i=s_j=0} \\
& \quad\quad \quad =  \sum_{\pi} (f_{k_1},P g_{k_{\pi 1}}) \cdots (f_{k_n},P g_{k_{\pi n}}),
\end{split}
\end{equation}
where the sum is over all permutations $\pi$ of $\{1,\dots,n\}$.

From Equ.~\eqref{largeYgen} and using the above results, we can now obtain any component of the coarse-grained metric. For instance, in distinct modes $k_1, \dots, k_n$, if we denote the components of the Kubo-Mori metric (given in Equ.~\eqref{zerotempapprox}) by $K^{k_1,\dots,k_n}$, then the components of the coarse-grained metric in our approximation are
\begin{equation}
\label{cgmetricomps}
K^{k_1,\dots,k_n} \bigotimes_{j=1}^n \begin{pmatrix} u_{k_j}^{-2}   & 0 \\ 0 & v_{k_j}^{-2}  \end{pmatrix}  K^{k_1,\dots,k_n}\,e^{-\sum_j k_j^2 \sigma^2}.
\end{equation}

We consider first the example of the mass term in the Hamiltonian, namely
\begin{equation}
\tilde V_2 = \frac 1 2 \int \tilde \phi(x)^2 dx - c \one 
\end{equation}
where $c \in \mathbb R$ is there so that $\ave{\tilde V_2}_\rho = 0$. In terms of the variables $\phi_k$, this is
\( 
V_2 = \frac 1 2 \sum_k \phi_k \phi_{-k} - c' \one. 
\)
In terms of the generating operator $G_f^A$, this is just
\begin{equation}
V_2 = - \frac 1 2 \sum_k \frac{\partial^2}{\partial s \partial t} G_{t \phi_k + s \phi_{-k}}^A |_{t=s=0}.
\end{equation}
Indeed, the identity component is automatically absent from this expression due to the fact that polynomials of different orders generated by $G_f^A$ are automatically orthogonal in terms of the Kubo-Mori metric, and the fact that $\ave{V,\one}_\rho^S = \ave{V}_\rho$.

Let's use the shorthand $P_s^t := (R^A_t)^\dagger P R^A_s$, where $P$ is given by Equ.~\eqref{largeYP}.
By applying the above differentiation to both $f$ and $g$ in Equ.~\eqref{largeYgen}, and using Equ.~\eqref{derivative}, and the fact that $(\phi_k,P_s^t \phi_{k'}) = 0$ if $k \neq k'$, we obtain (still for $y_\phi y_\pi \gg 1$),
\begin{equation}
\label{DV2} 
\begin{split}
\dist{V_2} &\simeq \frac 1 2 \sum_k \int ds \,dt\, (\phi_{k},P_s^t \phi_{k})  (\phi_{-k},P_s^t \phi_{-k}) \\
&= \frac 1 2 \sum_k \int ds\, dt\, (\phi_{k} \otimes \phi_{-k} , (P_s^t \otimes P_s^t ) \phi_{k} \otimes \phi_{-k} ).
\end{split}
\end{equation}
Using Equ.~\eqref{zerotempapprox}, each term of the sum can be compute explicitly in the zero temperature limit. The distinguishability density of $\tilde V_2$, which we denote
\begin{equation}
\label{distdens}
d(\tilde V_2) = \lim_{L\rightarrow \infty} L^{-d} \dist{V_2}
\end{equation}
is then obtain simply by replacing the sum by an integral in Equ.~\eqref{DV2}, which is
\begin{equation}
\label{dV2}
\begin{split} 
d(\tilde V_2) 
&\simeq \frac 1 {2^5}\int dk \, \frac{e^{-2 \sigma^2 k^2}}{\omega_{k}^2} \left[{ \frac{1}{u_{k}^4 \omega_{k}^4 } + \frac{1}{v_{k}^4 } }\right],
\end{split}
\end{equation}
where $u_k$ and $v_k$ are given by Equ.~\eqref{uk} and Equ.~\eqref{vk}, but in the limit $\beta \rightarrow \infty$, i.e., with $\coth \bigl( \tfrac{\beta \omega_k}2 \bigr) \rightarrow 1$.

We also want to consider the quartic interaction term 
\begin{equation}
\label{phi4def}
\begin{split}
\tilde V_4 &= \frac 1 {4!}\int \tilde \phi(x)^4 dx - \frac 1 2\mu^2 \int \tilde \phi(x)^2 dx - c \one, \\
\end{split}
\end{equation}
where the identity component compensates the operators' non-zero expectation value, and the second term is a counter term obtained from renormalisation (only to first order in perturbation theory). It is needed because the quartic term alone yields expectation values which diverge in the continuum limit (when an ultraviolet cutoff is removed), and hence does not by itself constitute a valid perturbation of the Hamiltonian in the continuum.

An ultraviolet regularisation is given simply by
\begin{equation}
\begin{split}
\tilde V_4 &= \frac 1 {4!} \int_\Sigma dx \, \frac{\partial^4}{\partial t^4} G^A_{t \tilde \phi(x)} |_{t=0}.\\
\end{split}
\end{equation}
This is due to the orthogonality of the polynomials generated by $G_f^A$ in terms of the Kubo-Mori metric, which guarantees that adding $\tilde V_4$ to the Hamiltonian has no influence on moments lower than four (to first order in the coupling constant). In particular, this means that $\tilde V_4$ does not influence the value of the ``macroscopic'' mass that is defined from the second moment in $\phi(x)$, which is the role played by $\mu$ in Equ.~\eqref{phi4def}.

To proceed, we need to express this interaction in term of our variables $\phi_k$. Observe that $\int \tilde \phi(x)^4 dx = \int dk_1\dots dk_4\, \delta(\sum_i k_i) \,\tilde \phi_{k_1} \dots \tilde \phi_{k_4}$. We use the infrared-regulated operator
\begin{equation}
V_4 = \frac{1}{4!} \sum_{k_1,\dots,k_4} \delta(\Sigma_i k_i) \,\frac{\partial^4 G^A_{\Sigma_i t_i \phi_{k_i}}}{\partial{t_1}\cdots \partial{t_4}} 
\end{equation}
so that
\(
\tilde V_4 = \lim_{L\rightarrow \infty} L^{d} V_4.
\)
Applying these derivatives to Equ.~\eqref{largeYgen} like in the previous example, we obtain (still for $y_\phi y_\pi \gg 1$),
\begin{equation}
\dist{V_4} \simeq \frac 1 {4!} \sum_{k_1\dots k_4}\int ds\, dt \,\delta(\Sigma_i k_i) \prod_{i=1}^4 (\phi_{k_i},P_s^t \phi_{k_i}),
\end{equation}
which yields at zero temperature
\begin{equation}
\label{distphi4}
\begin{split}
&d(\tilde V_4) \simeq \frac 1 {4! \,2^6}\int dk_1 \cdots dk_4 \, \delta(\Sigma_i k_i)\\
& \quad \times \frac{e^{-\sigma^2 \Sigma_i k_i^2}}{(\Sigma_i \omega_{k_i})^2} 
\Bigl[ \frac{1}{\prod_i v_{k_i}^2} + \frac{1}{\prod_i u_{k_i}^2 \omega_{k_i}^2}	\\
& \quad \quad + \sum_{\pi} \frac{1}{u_{k_{\pi(1)}}^2 u_{k_{\pi(2)}}^2 v_{k_{\pi(3)}}^2 v_{k_{\pi(4)}}^2  \omega_{k_{\pi(1)}}^2 \omega_{k_{\pi(2)}}^2  } \Bigr],
\end{split}
\end{equation}
where the products are over $i=1,\dots,4$ and the last sum is over all six permutations $\pi$ of the set $\{1,2,3,4\}$ which are such that $\pi(1) < \pi(2)$,  
and, of course, $u_k$ and $v_k$ are taken at $\beta = \infty$.

\subsection{Classical version}

For comparison, we also consider the same calculation but for the classical Klein-Gordon field, that is, 
\begin{equation}
A_k = \begin{pmatrix} \frac 1 {\beta \omega_k^2} & 0 \\ 0 & \frac 1 \beta \end{pmatrix}.
\end{equation}
The operator $P$ from Equ.~\eqref{fisherP} is $P = \oplus_k P_k$ with
\begin{equation}
P_k = A_k^2 (A_k + Y_k e^{k^2 \sigma^2} )^{-1}.
\end{equation}
In particular,
\begin{equation}
(\phi_k, P \phi_k) = \frac 1 {\beta \omega_k^2 + \beta^2 \omega_k^4 y_\phi^2 e^{k^2 \sigma^2}}.
\end{equation}
We obtain
\begin{equation}
d(\tilde V_2) = \frac 1 2 \int dk \,(\phi_k, P \phi_k)^2,
\end{equation}
and 
\begin{equation} 
d(\tilde V_4) = \frac{1}{4!} \int dk_1 \cdots dk_4 \,\delta(\Sigma_i k_i) \,\prod_{i=1}^4 (\phi_{k_i}, P \phi_{k_i}).
\end{equation}
This is the value of the ``basketball'' Feynman diagram, albeit for a modified propagator. A partially analytical solution for the standard propagator can be found in Ref.~\cite{Andersen2000}. However, one can use a more versatile numerical method to evaluate it, such as explained in Section~\ref{relevance}.

\subsection{Comparison with regulated Bures metric}

Other authors have considered the raw Bures metric dependent on a sharp momentum cutoff, e.g., in Ref.~\cite{Gu2010,Miyaji2015}. This is equivalent to using a coarse-graining channel $\mc E$ defined by $Y=0$ and $X$ the projector onto modes $|k| \le 1/\sigma$ for some scale $\sigma$. Technically, the target algebra of $\mc E$ should not contain any mode larger than $1/\sigma$, or this would violate Equ.~\ref{chancond}, hence $X$ is actually an isometry (satisfying $X^\dagger X = \one$). The resulting channel $\mc E$ then simply performs a partial trace over all modes $|k| > 1/\sigma$.

In this section, we want to examine under what conditions such a quantity would match, possibly approximately, the coarse-grained metrics we computed. 

The ``raw'' distinguishability of a Hamiltonian perturbation in the Bures metric $\ave{X,Y}_\rho = \tr(X \Omega_\rho^{-1}(Y))$ is defined by
\begin{equation}
\Omega_\rho(A) = \frac 1 2 (A \rho + \rho A).
\end{equation}
In the case where $\rho$ is pure: $\rho = \proj{\Omega_0}$, we can use Equ.~\eqref{metricatzero} to relate it to the Kubo-Mori metric as
\begin{equation}
\lim_{\beta\rightarrow \infty} \beta \ave{G_f^A,G_g^A}_\rho^S = 2 \ave{H^{-1} G_f^A + G_f^A H^{-1}, G_g^A }_\rho.
\end{equation}
What we want to compute is simply our coarse-grained distinguishability when $\mc E = {\rm id}$. Note that this does not imply $\mc R_\rho = {\rm id}$ because the ``source'' metric (Kubo-Mori) is different from the target one (Bures). Instead, we have
\begin{equation}
\begin{split}
\lim_{\beta  \rightarrow \infty}\beta \ave{\mc R(G_f^A), G_g^A}_\rho 
&= \lim_{\beta  \rightarrow \infty}\beta \ave{G_f^A,G_g^A}^S_\rho\\
&= 2 \ave{H^{-1} G_f^A + G_f^A H^{-1}, G_g^A }_\rho.
\end{split}
\end{equation}
Hence,
\begin{equation}
\mc R(G_f^A) = \frac 2 \beta ( G_f^A H^{-1} + H^{-1} G_f^A ) + \mc O((1/\beta)^0).
\end{equation}
Using also Equ.~\eqref{metricatzero}, the distinguishability is then
\begin{equation}
\begin{split}
&\ave{G_f^A,\mc E \mc R(G_g^A)}_\rho^S = \lim_{\beta \rightarrow 0} 2 \beta \ave{G_f^A, G_g^A H^{-1} + H^{-1} G_g^A }_\rho^S\\
&\quad\quad = 2 \,\rho((G_f^A)^\dagger H^{-2}G_g^A) + 2 \,\rho(G_g^A H^{-2}(G_f^A)^\dagger ). 
\end{split}
\end{equation}
Hence components of this metric in distinct modes $k_1, \dots, k_n$ are given by the matrix
\begin{equation}
\frac{2}{\bigl( \sum_i \omega_{k_i} \bigr)^2} \Bigl[ \bigotimes_{i=1}^n ( A_{k_i} + \frac i 2 \Delta) + \bigotimes_{i=1}^n ( A_{k_i} - \frac i 2 \Delta) \Bigr].
\end{equation}
In order to compare to the components of the coarse-grained metric in Equ.~\eqref{cgmetricomps}, observe that, for the Klein-Gordon field state at $\beta = \infty$,
\begin{align}
(A_k \pm \frac i 2 \Delta) A_k^{-1} (A_k \mp \frac i 2 \Delta) &= 0 \\
(A_k \pm \frac i 2 \Delta) A_k^{-1} (A_k \pm \frac i 2 \Delta) &= 2 (A_k \pm \frac i 2 \Delta).
\end{align} 

Hence, the raw Bures metric is given by the substitution $u_k^2 \rightarrow \tfrac 1 2 \omega_k^{-1}$, $v_k^2 \rightarrow  \tfrac 1 2  \omega_k$ and $X \rightarrow  \one$ in Equ.~\eqref{cgmetricomps}.

This suggests that, for the Klein-Gordon example, our coarse-grained metric essentially matches the raw metric (with a momentum cutoff) at zero temperature when $u_k^2 \ll \omega_k^{-1}$ and $v_k^2 \ll \omega_k$. This would violate the uncertainty condition $y_\phi y_\pi \gg 1$, but this condition is not fundamental as it can be easily alleviated by making $y_\phi$ and $y_\pi$ dependent on $k$.

\section{Applications}
\label{applications}

\subsection{Wilsonian relevance}
\label{relevance}

An interacting QFT comes with a regulator such as a momentum cutoff. In general, even if it may loosely correspond to the energy beyond which the theory loses validity, it is not considered to be physical, or at least its value is not known. As such, the predictions of the theory at low energies must be independent of it. In general, the parameters of the theory (entering the Hamiltonian) must be made dependent on the cutoff so as to yield constant predictions. This function of the Hamiltonian on the cutoff is the {\em renormalisation group} (RG). 

Wilson argued that the RG can also be reinterpreted in a different way~\cite{Wilson1975}. As the energy cutoff is lowered, the theory may simplify in the sense that some ``coupling constant'' in the Hamiltonian, such as the parameter $\lambda$ in $H + \lambda V$, tend to zero. Wilson interprets this as meaning that the term $V$ is not needed to make correct predictions at low energies. Hence the simpler theory defined by the Hamiltonian $H$ is a good effective description of the system at low energies. 

A more detailed way of addressing the question of whether $V$ is detectable under certain experimental constraints is of course the calculation of an actual statistical distinguishability between $H$ and $H + \lambda V$ as a function of explicit resolution parameters.

Specifically, the quantity $d(V)$ defined in Equ.~\eqref{distdens}, tells us how easy it is to distinguish a Gibbs state for $H$ compared to one for $H + \epsilon V$, to lowest order in $\epsilon$, as a function of the various resolutions $\sigma, y_\pi, y_\phi$, per unit of volume. The behaviour of $d(V)$ as a function of the ``physical'' resolution parameter $\sigma$ should tell us directly how relevant $V$ is: if it decreases as $\sigma$ increases then one may deem it ``irrelevant''.
But this doesn't quite work, because, since all distinguishability measures are contractive, $d(V)$ always decreases as $\sigma$ increases. However, it is a density: it corresponds to the effective distinguishability of the term $V$ for an observer having access only to some fine volume. When the resolution $\sigma$ gets worse, the observer has effectively access to fewer ``pixels'' per unit of volume, which is why $d(V)$ decreases. 

To meet the standard concept of relevance, we should rather ask how the distinguishability of $V$ depends on $\sigma$ if the observer still has access to the same volume {\em relative to} $\sigma$. This means considering instead the unitless quantity 
\begin{equation}
\delta(V) := \sigma^d d(V),
\end{equation}
which can now increase or decrease. 
This extra factor $\sigma^d$ serves the same purpose as the active rescaling of the fields in Wilson's approach to renormalisation. 

More specifically, we are interested in the polynomial behaviour of $\delta(V)$ as a function of $\sigma$. Hence we can directly extract the degree of the polynomial as 
\begin{equation}
\alpha(V) := \frac{\partial \log \delta(V)}{\partial \log \sigma}.
\end{equation}

For instance, consider the case where $y_\phi$ is small, which also requires $y_\pi$ large given the uncertainty relations. We have, roughly,
\begin{equation}
d(\tilde V_2) \propto \int_{|k|<1/\sigma} dk \,\frac{1}{\omega_{k}^4} 
\end{equation}
Therefore, for $\sigma \ll \tfrac 1 m$ (or a massless theory), $d(\tilde V_2) = \mc O(\sigma^{4-d})$, and for $\sigma \gg \tfrac 1 m$, $d(\tilde V_2) = \mc O(\sigma^{-d})$. This yields $\alpha(\tilde V_2) \simeq 4$ below the mass scale and $\alpha(\tilde V_2) \simeq 0$ above the mass scale.

This can be understood as follows: as the ``pixel size'' $\sigma$ increases, more information about the mass is gained per pixel, until they reach the mass scale (which is also the correlation length), at which point no more new information is gained. Hence Information about the mass is found in large scale features of the state.

This corresponds to the Wilsonian analysis which would say that $\tilde V_2$ is {\em relevant} for a massless theory, and marginal in a massive theory.  

With a large field uncertainty $y_\phi$, the scaling is qualitatively different. Indeed, below the mass scale,
\begin{equation}
d(\tilde V_2) \propto \int_{|k|<1/\sigma} dk \frac{1}{y_\phi^4 \omega_{k}^6}, 
\end{equation}
which yields $\alpha(\tilde V_2) \simeq 6$.

As a further illustration, we want to compute the distinguishability scaling $\alpha(\tilde V_4)$ for the $\phi^4$ interaction term from Equ.~\eqref{distphi4}. 
We can make a similar ``back of the envelop'' analysis for the behaviour of $d(\tilde V_4)$ for $y_\pi$ very large and $y_\phi$ vanishingly small. This is
\begin{equation}
\label{dV4backfoenvelop}
d(\tilde V_4) \propto \int_{\sum_i k_i^2 < \frac 1 {\sigma^2}} \,\frac{\delta(\Sigma_i k_i)\, dk_1 \cdots dk_4 }{\bigl( \Sigma_i \omega_{k_i} \bigr)^2 \omega_{k_1} \cdots \omega_{k_4}}.
\end{equation}
At $m=0$, $\omega_k = k$. Counting the powers of $k$ inside the integral yields the guess $\alpha(\tilde V_4) \simeq 2(3-d)$ for $ 2 \le d \le 6$, which can be verified using the numerical method explained below. Hence we recover the expected marginal dimension $d_0 = 3$ above which the $\phi^4$ interaction is irrelevant at large scales. However, various other results can be obtained depending on the field resolutions $y_\phi$ and $y_\pi$, as can be anticipated by the resulting different powers in $\omega_k$ in Equ.~\eqref{dV2}.
 
In order to obtain an efficient numerical method to evaluate the integrals, consider that, if 
\begin{equation} 
d(V) = \int dK f(K),
\end{equation}
where $K=(k_1,\dots,k_n)$, then
\begin{equation}
\begin{split}
\alpha(V) &= \frac{\partial \log d(V)}{\partial \log \sigma} \\
&= \sigma \frac 1 {d(V)} \int dK f(K) \frac{\partial \log f(K)}{\partial \sigma}\\
&=  \sigma \,\Bigl\langle \frac{\partial \log f(K)}{\partial \sigma} \Bigr\rangle_f.\\
\end{split}
\end{equation}
The last term involves an expectation value in terms of the probability density $K \mapsto f(K)/\int dK f(K)$. Defining 
\begin{equation}
H(K) = -\log f(K),
\end{equation}
the integral can be evaluated using the metropolis algorithm for the classical Hamiltonian $K \mapsto H(K)$ at temperature $1$. 

\begin{figure} 
\ifnjp
\begin{center}
\includegraphics[width=0.7\columnwidth]{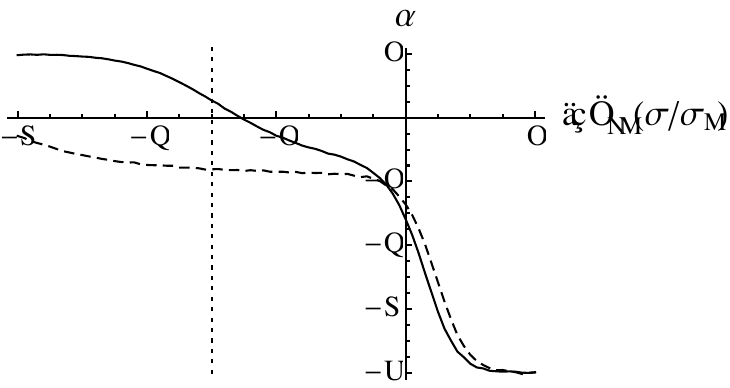}
\end{center}
\else
\includegraphics[width=0.9\columnwidth]{gaussian_figure1.pdf}
\fi
\caption{Scaling exponent $\alpha$ for the distinguishability of the $\phi^4$ interaction as a function of the scale (resolution) $\sigma$, where $\sigma_0 := 1/m$ is the correlation length. For both curves, $y_\pi^2 = 10^{10} / \sigma_0$, and the dimension of space is $d=4$. The solid curve is for $y_\phi^2 = 10^{-3} \sigma_0$ (a valued marked on the scale axis by the vertical dotted line). For comparison, the dashed curve is for the much smaller $y_\phi^2 = 10^{-7} \sigma_0$, for which $\alpha$ behaves essentially as predicted by Equ.~\eqref{dV4backfoenvelop}, i.e., $\alpha \simeq -2$ below the mass scale.}
\label{fig1}
\end{figure}

If we keep $y_\phi$ and $y_\pi$ independent of $\sigma$, those values introduce two lengthscales in additions to the mass, at which the behaviour of $d(\tilde V_4)$ changes. The effect of an increase in $y_\phi$ is shown in Figure~\ref{fig1}. In this example with four spatial dimensions ($d=4$), an increased value of $y_\phi$ renders the interaction $\tilde V_4$ relevant ($\alpha>0$) instead of irrelevant ($\alpha < 0$) below the mass scale for $\sigma < y_\phi^2$.

\subsection{Dimensionality reduction}

An approach to renormalisation proposed in Ref.~\cite{Beny2013,Beny2015a} consists in determining the eigenvalues and eigenvectors for the linear map $\mc E \mc R_\rho$, which is self-adjoint with respect to the source metric. The original interpretation of the calculation was different, because $\mc R_\rho$ (which was then called $\mc R_\rho^\dagger$) occurred geometrically as the pull-back induced by $\mc E$ on cotangent spaces to states (we refer to Ref.~\cite{Beny2015a} for details). However, technically the calculations are identical.  

In that approach, the eigenvectors are interpreted as observables, and the eigenvalues (which are always in the interval $[0,1]$) tell us how much these observables lose ``distinguishability'' under coarse-graining. A smaller effective tangent space (to states) is given by ignoring those observables which lose too much distinguishability, i.e., which are not ``relevant'' enough, by deeming two tangent vectors effectively identical if they give the same expectations values for the more relevant observables. 

For the Klein-Gordon field example, the components of $\mc E \mc R_\rho $ as a linear map can be determined using the observations made at the end of Section~\ref{distinggeneral}, namely that $ \mc E \mc R_\rho$ is block-diagonal, with each block associated with momentum modes $k_1,\dots,k_n$. For $y_\pi y_\phi \gg 1$ and at zero temperature, one can directly see from Equ.~\eqref{cgmetricomps} that the components of one block are given by the matrix
\begin{equation}
\label{ERblock}
\bigotimes_{j=1}^n \begin{pmatrix} u_{k_j}^{-2}   & 0 \\ 0 & v_{k_j}^{-2}  \end{pmatrix}  K^{k_1,\dots,k_n}\,e^{-\sum_j k_j^2 \sigma^2}
\end{equation}
where the matrices $K^{k_1,\dots,k_n}$ are equal to the right hand side of Equ.~\eqref{zerotempapprox}. 

For instance, for $n=1$ one recovers the single-mode result presented in Ref.~\cite{Beny2015a} (at $\beta \rightarrow \infty$). For $n=2$, and $k_1=-k_2=k$ for simplicity, we obtain the eigenvalue
\[
\eta_1 = \frac{e^{-2 k^2 \sigma^2}}{2 \beta \omega_k u_k^2 v_k^2} + \mc O(\beta^{-2})
\]
with corresponding eigenvector (observable)
\[
A_1 = \phi_k \pi_{-k} + \pi_{-k} \phi_k
\]
and
\[
\eta_2 = \frac{e^{-2 k^2 \sigma^2}}{4 \beta} \Bigl[ \frac{1}{\omega_k^3 u_k^4 4} + \frac{\omega_k}{v_k^4} \Bigr] + \mc O(\beta^{-2}),
\]
for
\[
A_2 = u_k^4 \omega_k^2 |\pi_k|^2 - v_k^4 |\phi_k|^2.
\]
The other two eigenvalues are of order $\beta^{-2}$. Two orthogonal polynomials spanning that space are given by 
\[
A_3 = i( \phi_k \pi_{-k} - \pi_{-k} \phi_k)  \; \text{ and } \; A_4 = \omega_k^2 |\phi_k| ^2 + |\pi_k|^2. 
\]

\section{Outlook}
\label{outlook}

We developed a practical way of computing the effective distinguishability between Hamiltonian perturbations, taking experimental limitations explicitly into accounts. These resolutions parameters are needed to obtain a finite measure of distinguishability.
We developed a calculation procedure applicable directly to Gaussian states (free fields), to arbitrary perturbations, and for all contractive metrics. Moreover, we showed in the context of the Klein-Gordon field that all contractive metrics give the same result provided that the phase-space resolution is much coarser than $\hbar$.

Although the quantities computed, namely the component of a metric, only give the distinguishability between a state and a perturbation of the state to first order in the coupling constant, this can be used in principle to compute geodesic distances between any two state (or at least upper bounds to it). This is where the geometric nature of the formalism becomes useful. For instance, the distinguishability between a state $\rho_{AB}$ on two systems $A$ and $B$ and the product state $\rho_A \otimes \rho_B$ (where $\rho_A$ and $\rho_B$ are reduced states of $\rho_{AB}$) is a genuine measure of entanglement when $\rho_{AB}$ is pure, such as a ground state. This can be used to compute an upper bound to the ground state entanglement on any region $A$. For instance, if $\rho$ is the ground state of a local field Hamiltonian, then a path to the product state $\rho_A \otimes \rho_{A^C}$ may be obtained by progressively increasing the field's mass on the boundary. In the purely Gaussian case, the recent results in Ref.~\cite{Banchi2015} may provide a more direct route. However, our method can in principle be extended perturbatively beyond Gaussian states using the approach proposed in Ref.~\cite{Beny2015a} in the classical case. Namely, a perturbation of a Gaussian state $\rho$ yields a perturbation of the linear map $\mc E \mc R_\rho$, whose components can in principle be computed to each order. 

A generalisation to fermionic fields should also be possible using the fermionic Gaussian formalism, such as developed in Ref.~\cite{Bravyi2004}.

\vspace{0.5cm}
\section*{Acknowledgements}

The author is grateful to Tobias Osborne for discussions related to this work and the suggestion to consider perturbation theory to simplify the problem in the zero temperature limit. 
This work was supported by the ERC grants QFTCMPS and SIQS and by the cluster of excellence EXC 201 Quantum Engineering and Space-Time Research.

\ifnjp
\section*{References}
\bibliographystyle{iopart-num}
\fi

\ifquantum
\bibliographystyle{plainnat}
\fi

\bibliography{../complete.bib}

\end{document}